\documentclass[reprint,prb,superscriptaddress,showpacs]{revtex4-2}%

\usepackage{xr-hyper}
\setcitestyle{maxcitenames=6}
\usepackage{gensymb}
\usepackage{graphicx}
\usepackage{xcolor}
\usepackage{amsmath}
\usepackage{amsfonts}
\usepackage{amssymb}
\usepackage{framed}
\usepackage{float}
\usepackage[normalem]{ulem}
\usepackage{lipsum}
\usepackage[mathlines,switch]{lineno}
\providecommand{\U}[1]{\protect\rule{.1in}{.1in}}
\makeatletter
\renewcommand*{\fnum@figure}{{\normalfont\bfseries \figurename~\thefigure}}
\renewcommand*{\@caption@fignum@sep}{\textbf{ : }}
\makeatother
\usepackage{hyperref}
\hypersetup{
    colorlinks=true,
    linkcolor=blue,
    filecolor=magenta,      
    urlcolor=cyan,
}

\usepackage{siunitx}
\usepackage{adjustbox}
\usepackage{tabularx}
\usepackage[table]{xcolor}
\usepackage{booktabs}
\usepackage{gensymb}
\usepackage{xcite}

\makeatletter

\usepackage[normalem]{ulem}

\makeatother

\usepackage{hyperref}
\hypersetup{
    colorlinks=true,
    linkcolor=blue,
    filecolor=magenta,      
    urlcolor=cyan,
}
\usepackage{hyperref}
\usepackage{mhchem}

\begin{document}

%\preprint{APS/123-QED}

\title{Learning the Geometry of Materials Space with Trained Neural Tangent Kernels}% Force line breaks with \\

\author{Daniel Kaplan}
\email{daniel.kaplan@ucf.edu}
\affiliation{Center for Materials Theory, Department of Physics and Astronomy, Rutgers University, Piscataway 08854 New Jersey USA}
\affiliation{Department of Physics and Astronomy, Rutgers University, Piscataway 08854 New Jersey USA}
%\affiliation{Department of Physics, University of Pennsylvania, Philadelphia, Pennsylvania 19104, USA}
\affiliation{Department of Physics, University of Central Florida, Orlando, Florida 32816, USA}
\affiliation{NanoScience and Technology Center, University of Central Florida, Orlando, Florida 32816, USA}
\author{Gabriel Kotliar}
\affiliation{Center for Materials Theory, Department of Physics and Astronomy, Rutgers University, Piscataway 08854 New Jersey USA}
\affiliation{Department of Physics and Astronomy, Rutgers University, Piscataway 08854 New Jersey USA}
 \author{Anirvan M Sengupta}
\affiliation{Center for Computational Quantum Physics, Flatiron Institute, New York, NY 10010}
\affiliation{Center for Computational Mathematics, Flatiron Institute, New York, NY 10010}
\affiliation{Department of Physics and Astronomy, Rutgers University, Piscataway 08854 New Jersey USA}
\date{\today}

\begin{abstract}
Neural networks trained to predict materials properties learn representations whose physical content can be difficult to extract. We use the neural tangent kernel (NTK), evaluated at trained parameters, to recover the geometry learned by a network that predicts superconducting transition temperatures from elemental composition alone. Kernel ridge regression with the trained NTK closely reproduces the network's predictive performance and substantially outperforms regression with the initial kernel. Spectral analysis and multidimensional scaling reveal a geometry dominated by a small number of directions. Clustering and local neighborhood analysis organize compounds into recognizable superconducting families and connect materials with related chemical and correlated-electron properties despite substantial differences in $T_c$. These results establish trained tangent kernels as practical probes of learned materials representations, connecting property prediction with the exploration of materials space.
\end{abstract}

\maketitle
\section{Introduction}
  \begin{figure*}[!t]
    \centering
     \includegraphics[width=1.0\textwidth]{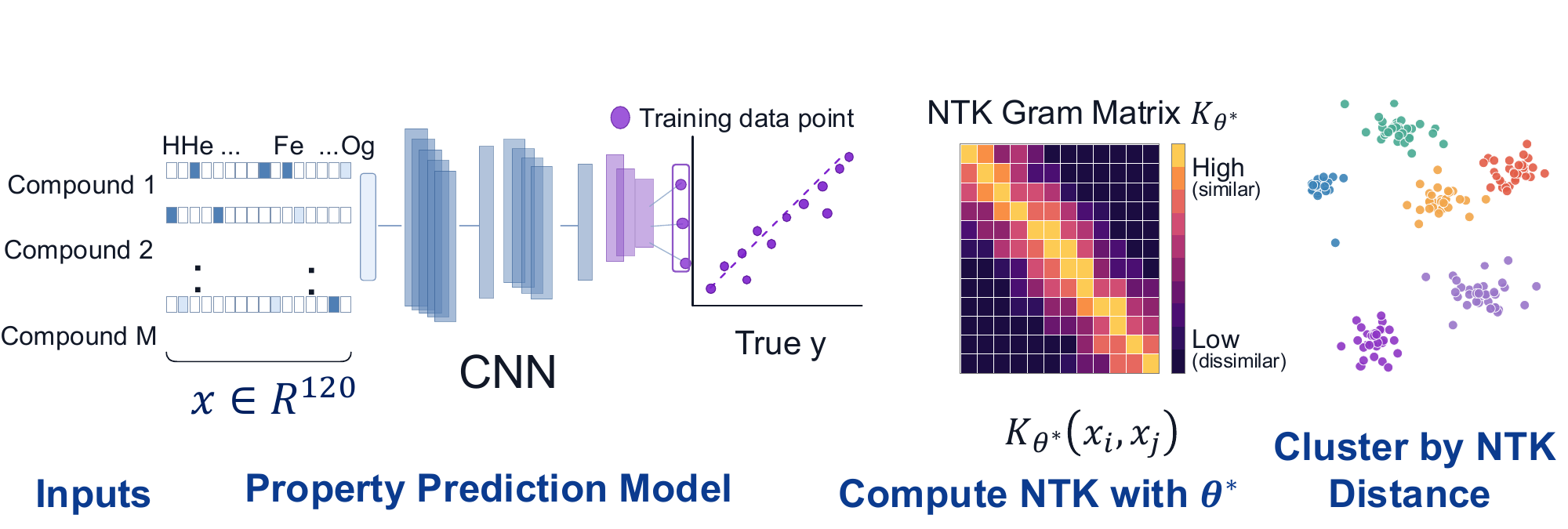}
   \caption{
    Schematic overview of the workflow. Minimal compositional
descriptors are used to train a neural network predictor for
superconducting properties. The neural tangent kernel (NTK)
evaluated at trained parameters induces a task-adaptive
similarity geometry on materials space.  Dimensional reduction and clustering of this geometry reveal semantically
 meaningful structure learned through supervised training.    
   }
    \label{fig:overview}
 \end{figure*}
Neural networks trained on materials datasets can learn representations that
support accurate predictions for unseen compounds~\cite{Reiseretal2022,Moosavietal2020}.
The reuse of learned features across related tasks underlies transfer learning,
which has become an important tool in materials
science~\cite{yosinski2014transferable,buterez2024,rocken2025}.
Understanding the physical content of these representations remains a central
challenge in scientific machine learning~\cite{Arzanietal2023}. Learning meaningful representations is a long-standing goal in machine
learning. Convolutional networks learn hierarchical
features~\cite{lecun1998gradient,he2016deep}, while independent component
analysis and sparse coding identify statistical structure directly from
data~\cite{bell1997independent,olshausen1996emergence}. These approaches helped
establish modern representation learning~\cite{bengio2013representation}.
In materials science, self-supervised learning, manifold learning, and
generative approaches increasingly exploit structural, graph-based, or
multimodal information to organize and explore materials
space~\cite{suzuki2022self,kim2026chartingemergentlowdimensionalmanifold,okabe2026structural,panigrahi2026graphlethistogramrepresentationdatabase}.

Our earlier work~\cite{kaplan2025deep} demonstrated that competitive prediction
of superconducting transition temperatures is possible using elemental
composition alone, without engineered descriptors such as
Magpie~\cite{ward2016general,stanev2018machine} or the crystal structures used
by many graph-based materials
representations~\cite{xie2018crystal,choudhary2021atomistic}.
The success of this predictor raises a complementary question: what physical
relationships between materials has the network acquired through supervised
training, and how can they be recovered? Here we extract and interpret this
learned representation from the trained predictor itself, without introducing
additional descriptors or auxiliary representation-learning objectives.

We use the neural tangent kernel (NTK) evaluated at the trained parameters
as a probe of this representation. Originally introduced to analyze
optimization in wide neural networks~\cite{jacot2018neural}, the NTK can
evolve substantially during finite-width training and become aligned with
the learned task~\cite{fort2020deep,shan2021theory,canatar2022kernel}.
Its geometry therefore provides access to relationships acquired during
training, extending its use from the analysis of optimization to the
interpretation of learned materials representations.

For a scalar predictor $f(x,\theta)$ with trained parameters $\theta^*$,
the tangent feature map
$\phi(x)=\left.\nabla_\theta f(x,\theta)\right|_{\theta=\theta^*}$
defines the kernel $K(x,x')=\phi(x)^{\mathsf T}\phi(x')$.
The corresponding distance,
$d(x,x')=\|\phi(x)-\phi(x')\|_2$,
compares the local responses of the predictions to parameter perturbations.
Compounds are close in this geometry when their predictions respond similarly
to small changes in the trained parameters, even when their compositional
vectors have little or no elemental overlap.

We show that kernel ridge regression with the trained NTK achieves predictive
performance comparable to that of the original network, while regression
with the initial kernel performs substantially worse. Spectral analysis and
multidimensional scaling reveal a geometry dominated by a small number of
directions. Applying \emph{unsupervised} clustering to this embedding
identifies recognizable regions of materials space, including cuprates,
iron-based superconductors, and non-superconducting compounds. The cluster
names are assigned after clustering, allowing us to interpret the physical
organization encoded in the learned geometry.

Local neighborhoods reveal further relationships between chemically distinct
materials. A particularly striking example is the proximity of CeCoIn$_5$
and YbRh$_2$Si$_2$: these compounds share no constituent elements, yet both
exhibit heavy-fermion behavior and are central to the study of correlated
electron physics~\cite{gegenwart2008quantum,petrovic2001heavy,zhou2013visualizing}.
Their connection emerges without supplying this phenomenology as a training
label.

These neighborhoods extend the use of the predictor to identifying related
compounds and locating regions in which nearby materials span a broad range
of $T_c$. Such regions provide targets for further materials exploration.
The trained NTK thus turns an existing property predictor into a tool for
navigating materials space and provides a learned geometry for subsequent
transfer learning and materials discovery.
An overview of the workflow is shown in Fig.~\ref{fig:overview}.

\section{From Influence Functions to Trained Tangent Kernels}

A natural data-space definition of interpretability is influence: how would 
the prediction at one material change if another material in the training set
were infinitesimally reweighted. This question is especially useful in materials science, where interpretation often proceeds by identifying representative compounds, nearby families, and physically analogous systems.

Let \(z_i=(x_i,y_i)\) denote a training example, and $y_i$ the empirical ground truth. For the purposes of what follows, $x_i$ will be a vector of dimension $x \in \mathbb{R}^{120}$\footnote{The value $120$ here is meant to represent the number of elements in the periodic table, padded with two more to allow the factorization $12\times 10$ for an image input to the CNN.} and $y_i$ is a real number denoting $T_c$. For the empirical loss,
\begin{align}
\mathcal L(\theta)=\frac{1}{N}\sum_{i=1}^N \ell(f_\theta(x_i),y_i),
\end{align}
we arrive at \(\theta^\ast\) as the trained parameters. Following the influence-function
construction \cite{koh2017understanding}, we up-weight point \(i\) by an infinitesimal amount \(\epsilon\),
\begin{align}
\theta_{\epsilon,i}
=
\arg\min_\theta
\left[
\mathcal L(\theta)+\epsilon \ell(f_\theta(x_i),y_i)
\right].
\end{align}
To first order in \(\epsilon\),
\begin{align}
\left.
\frac{d\theta_{\epsilon,i}}{d\epsilon}
\right|_{\epsilon=0}
=
-H_{\theta^\ast}^{-1}
\nabla_\theta \ell_i(\theta^\ast),
\end{align}
where \(H_{\theta^\ast}=\nabla_\theta^2\mathcal L(\theta^\ast)\). The
corresponding change  in the prediction at another point \(x_j\) \cite{koh2017understanding} is
\begin{align}
\mathcal I_{i\to j}
=
-\nabla_\theta f_{\theta^\ast}(x_j)^T
H_{\theta^\ast}^{-1}
\nabla_\theta \ell_i(\theta^\ast).
\end{align}
For squared loss, with residual \(r_i=f_{\theta^\ast}(x_i)-y_i\), this becomes
\begin{align}
\mathcal I_{i\to j}
=
-r_i\,
\phi_j^T H_{\theta^\ast}^{-1}\phi_i,
\qquad
\phi_i=\nabla_\theta f_{\theta^\ast}(x_i).
\end{align}
Thus the influence function decomposes into a residual factor and a
Hessian measure for similarity between two data points. In our case, these data points are in fact different materials.

Often, as for our $f(x,\theta^{*})$, computing this expression exactly is prohibitive: 
the Hessian is large, potentially ill-conditioned, and expensive to invert.
We are therefore motivated to replace the full nonlinear model by its local linearization at
the trained parameters,
\begin{align}
f_{\theta^\ast+\delta\theta}(x)
\simeq
f_{\theta^\ast}(x)+\phi(x)^T\delta\theta.
\end{align}
Given the appearance of $\phi(x)^{T}$, it is useful to define the Gram object, $K_{ij} \phi_i^T\phi_j$, which constitutes a kernel in feature space.  The connection to influence becomes explicit under the Gauss-Newton
approximation \(H_{\theta^\ast}\simeq J^T J+\lambda I\), where \(J\) is the
Jacobian with rows \(\phi_i^T\). In this approximation, considering the Sherman-Morrison-Woodbury identity \cite{bernstein2018scalar},
\begin{align}
J(J^TJ+\lambda I)^{-1}J^T
=
K^\ast(K^\ast+\lambda I)^{-1}.
\end{align}
Consequently, the regularized influence of training point \(i\) on prediction
\(j\) is controlled by
\begin{align}
\mathcal I^\lambda_{i\to j}
\simeq
-r_i
\left[
K(K+\lambda I)^{-1}
\right]_{ji}.
\end{align}
Thus the trained NTK, being a
positive-semidefinite kernel, is an object whose spectral geometry determines the
regularized influence operator of the linearized network. In the
analysis below, we use \(K\) and distances derived from it as a stable,
label-independent proxy for this influence geometry.

\section{The Trained Neural Tangent Kernel and Representation Fidelity}
The trained NTK can be shown to serve functions which are broadly related to the tasks and modalities of the neural network itself.
For example, we first test the ability of the NTK for prediction. We carry out the following procedure and compare its performance with that of the original network. Let our inputs be $\bf{x}$ and corresponding labels be $\bm{y}$ (here, $T_c$). 

The predictive mean for a new test point $\bm{x}_*$ in this procedure would be given by:
\begin{equation}
    f(\bm{x}_*) =K(\bm{x}_*, \bm{X}) \left( K(\bm{X}, \bm{X}) + \lambda \bm{I} \right)^{-1} \bm{y}
\end{equation}
where $K(\bm{X}, \bm{X})$ is the $N \times N$ post-training Gram matrix,   $K(\bm{x}_*, \bm{X})$ is the $1 \times N$ cross-kernel vector between the test point and training data, $\lambda$ is the ridge regularization penalty, acting as a noise floor to prevent overfitting. Generally, this is known as the KRR (Kernel ridge regression) problem \cite{krige1951statistical}. 

KRR is solved in a two-step process utilizing dual coefficients. Instead of matrix inversion, we solve for the dual coefficient vector $\bm{\alpha}$:
\begin{equation}
    \left( K(\bm{X}, \bm{X}) + \lambda \bm{I} \right) \bm{\alpha} = \bm{y}
\end{equation}
Once $\bm{\alpha}$ is computed, inference for an arbitrary test set $\bm{X}_{\text{test}}$ becomes a highly efficient matrix-vector multiplication:
\begin{equation}
    f(\bm{X}_{\text{test}}) = K(\bm{X}_{\text{test}}, \bm{X}) \bm{\alpha}
\end{equation}

\begin{table}[htbp]
    \centering
    \caption{Comparison between NN results and Kernel Regression: $T_c$ is the prediction $T_c^0$, being the original data.}
    \label{tab:comparison_kernel}
    \resizebox{\columnwidth}{!}{
    \begin{tabular}{@{} l l l l @{}} 
        \toprule
        \textbf{Method} & \textbf{mean $|T_c-T_c^0|$} & \textbf{rms $(T_c-T_c^0)$} & \textbf{$\bar{T}_c$} \\
        \midrule
       Neural Network & 2.11K & 7.20K & 17.5K \\ 
       Kernel Regression (trained) & 2.24K & 7.61K & -- \\ 
       Kernel Regression (untrained) & 21.7K & 31.9K & -- \\
        \bottomrule
    \end{tabular}}
\end{table}
The comparable performance between the neural network and kernel ridge regression, shown in Table~\ref{tab:comparison_kernel}, suggests that a substantial fraction of the learned representation is encoded in the trained NTK.
In contrast, kernel regression using the NTK at random initialization performs substantially worse, indicating that the predictive structure captured by the trained NTK is not inherited from the architecture alone, but emerges during training through feature learning. This observation motivates treating the trained NTK not merely as a visualization tool, but as an interpretable representation of the learned predictor itself.

\section{Neural Tangent Kernel Distance and Feature Learning}

Given a trained neural network $f_\theta(x)$, we define the neural tangent kernel (NTK) at the trained parameters $\theta^*$ as
\begin{equation}
K(x,x') = \nabla_\theta f_\theta(x)^\top \nabla_\theta f_\theta(x') \Big|_{\theta=\theta^*}.
\end{equation}
This kernel can be interpreted as an inner product in a feature space $\phi(x) = \nabla_\theta f_\theta(x)$.
Based on this inner product, we define a distance:
\begin{equation}
d(x,x')^2 = K(x,x) + K(x',x') - 2K(x,x'),
\label{eq:euc}
\end{equation}
corresponding to Euclidean distance in the induced feature space. This geometry in the feature space is closely related to kernel PCA and related manifold learning methods, where low-dimensional structures are extracted from the spectral decomposition of similarity kernels \cite{scholkopf1998nonlinear,coifman2006diffusion}.

\begin{figure}
    \centering
    \includegraphics[width=0.9\columnwidth]{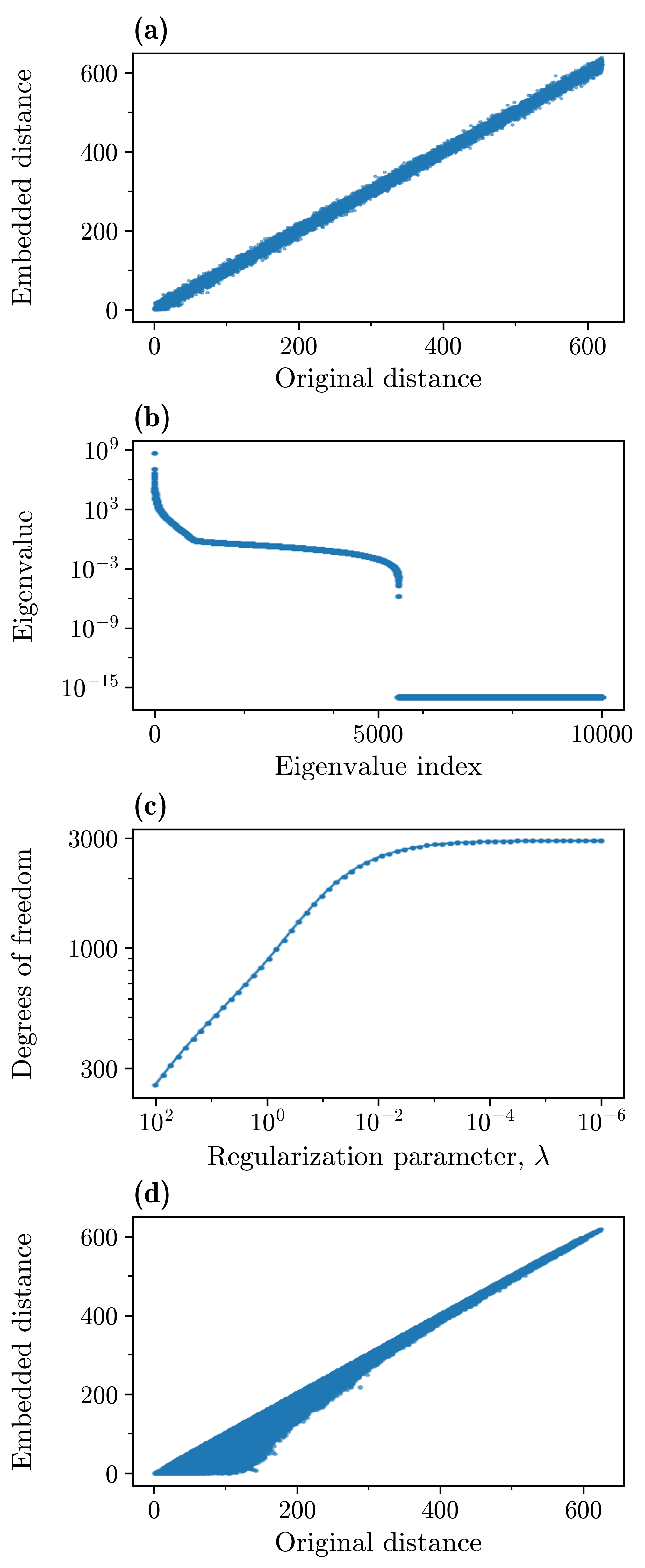}
    \caption{Properties and structure of the NTK. (a) The Shepard plot, i.e., embedding distance $||z_i-z_j||$ vs Euclidean distance of an untrained NTK realization, derived first the first $N=1$ step of training. No embedding is generated. (b) Eigenvalue spectrum of the train NTK. Approximately two principal eigenvalues can be discerned. (Note the log scale). (c) Kernel Ridge Regression (KRR) properties: rapid convergence of the KRR with respect to the regularization $\lambda$ confirms the relative number of useful degrees of freedom (non-null eigenvalues in (b)). (d) Shepard plot of the trained NTK, indicating a significantly different embedding of the data. }
    \label{fig:shepard}
\end{figure}

As Fig.~\ref{fig:shepard}(a) shows, the initial NTK largely preserves compositional distances inherited from the input representation. By contrast, as presented in Fig~\ref{fig:shepard}(d), the trained NTK substantially reorganizes the geometry, bringing many compositionally distant but related materials closer together in feature space. This suggests that training induces a compression of materials space aligned with superconducting properties.

Classical NTK theory considers regimes in which the kernel remains fixed during training \cite{jacot2018neural,du2019gradient}. However, in practical settings, the NTK evolves and becomes aligned with the target function \cite{fort2020deep,shan2021theory}. This reflects feature learning: the network adapts its representation so that relevant directions are amplified. As a result, the trained NTK often exhibits structured, low-rank spectra, with leading eigenspaces capturing task-relevant variation \cite{canatar2022kernel}. 

Having established that training substantially reorganizes the geometry induced by the NTK, we now investigate the dimensional structure of this learned representation through spectral analysis. Our sample space is composed of $N_s=16414$ data points, making the NTK an $N\times N$ object. The coordinates associated with the $x_i$s can be thought of as independent degrees of freedom. To assess their importance, we examine the effect of regularization $\lambda$ on the eigenspectrum of our problem. Conveniently, this corresponds to the estimate of the ridge regression on our kernel \cite{vovk2013kernel,chang2017distributed}. We let,
\begin{align}
    d_f=\textrm{Tr}(K(K+\lambda I)^{-1}),
\end{align}
where $d_f$ is the proxy for the true degrees of freedom in the kernel. In Fig.~\ref{fig:shepard}(c), we plot $d_f$ as a function of $\lambda$.

The trained kernel its spectral weight in a small number of leading directions. We quantify this concentration using the regularized effective dimension, $d_{\mathrm{eff}}(\lambda)=\sum_a\kappa_a/(\kappa_a+\lambda)$, and the fraction of centered feature variance retained by the leading MDS coordinates. These measures distinguish the dominant geometric directions from the larger set of nonzero kernel modes.

This can be further strengthened by computing the eigenvalues exactly in the sample space $N_s$. In Fig.~\ref{fig:shepard}(b) we show the characteristic eigenvalue spectrum of the NTK, with $N_s = 10000$ (for visual clarity, we truncate all zero eigenvalues beyond this point), on a log scale of eigenvalue intensity. We find that a select number of features $N_f < 10$ possess eigenvalues in the range of $10^6-10^9$. This reveals the \textit{most significant} features which the NTK is able to resolve without additional training. For this reason, this is a first glimpse into the possibility of extracting "transferable" learning and a low-dimensional manifold which learned through the supervised training.

To visualize the structure encoded in the trained NTK distances, we construct a low-dimensional embedding using classical multi-dimensional scaling (MDS) \cite{borg2005modern}. Given the pairwise distance matrix $d(x_i,x_j)$ defined in Eq. (5), classical MDS seeks coordinates $z_i \in \mathbb{R}^K$ whose Euclidean distances approximately preserve the original NTK geometry:
\begin{align}
||z_i-z_j||\approx d(x_i,x_j).
\end{align}
Equivalently, MDS constructs an embedding by diagonalizing the double-centered distance matrix. Concretely, let,
\begin{align}
\Pi=I-\frac1N\mathbf1\mathbf1^{\mathsf T},
~~
B=-\frac12\Pi D^{(2)}\Pi=\Pi K\Pi.
\end{align}
Here, $D^{(2)}$, 
producing a low-dimensional representation that preserves the dominant geometric relationships between our inputs, which are materials ($x_i$). 
Applying this to the NTK distances from the SuperCon dataset, we find that even for $K=2$, the explained variance is $92\%$, and going up to $K=3$ increases it marginally to $95.3\%$. This is consistent with the effective low-rank structure of the Kernel as implied by Figs.~\ref{fig:shepard}(b)-(c).
\section{NTK Geometry of Materials Space}

A central question, now, is whether this low-dimensional embedding
captures physically meaningful structure or merely reflects
a geometric compression of the data. To probe this question,
we examine both the clustering structure induced in the
embedded space and correlations with physically relevant
observables not used directly in training. The emergence of
coherent superconducting families together with correlations
between different chemical families implies that the NTK embedding encodes meaningful structure within materials space rather than arbitrary geometric
organization.

\subsection{Clustering the Latent Vectors}

We apply $k$-means \cite{mcqueen1967some} to the latent vectors $\mathbf{z}_i$. We partition the data into $C$ sets $\mathbf{S} = \{S_1, S_2, \dots, S_C\}$ by minimizing the within-cluster variance:
\begin{equation}
    \text{arg min}_{\mathbf{S}} \sum_{c=1}^C \sum_{\mathbf{z} \in S_c} \|\mathbf{z} - \boldsymbol{\mu}_c\|^2
    \label{cluster_size}
\end{equation}
where $\boldsymbol{\mu}_c$ is the centroid of cluster $S_c$.

\begin{figure*}[ht]
  \vskip 0.2in
  \begin{center}
    \centerline{\includegraphics[width=\textwidth]{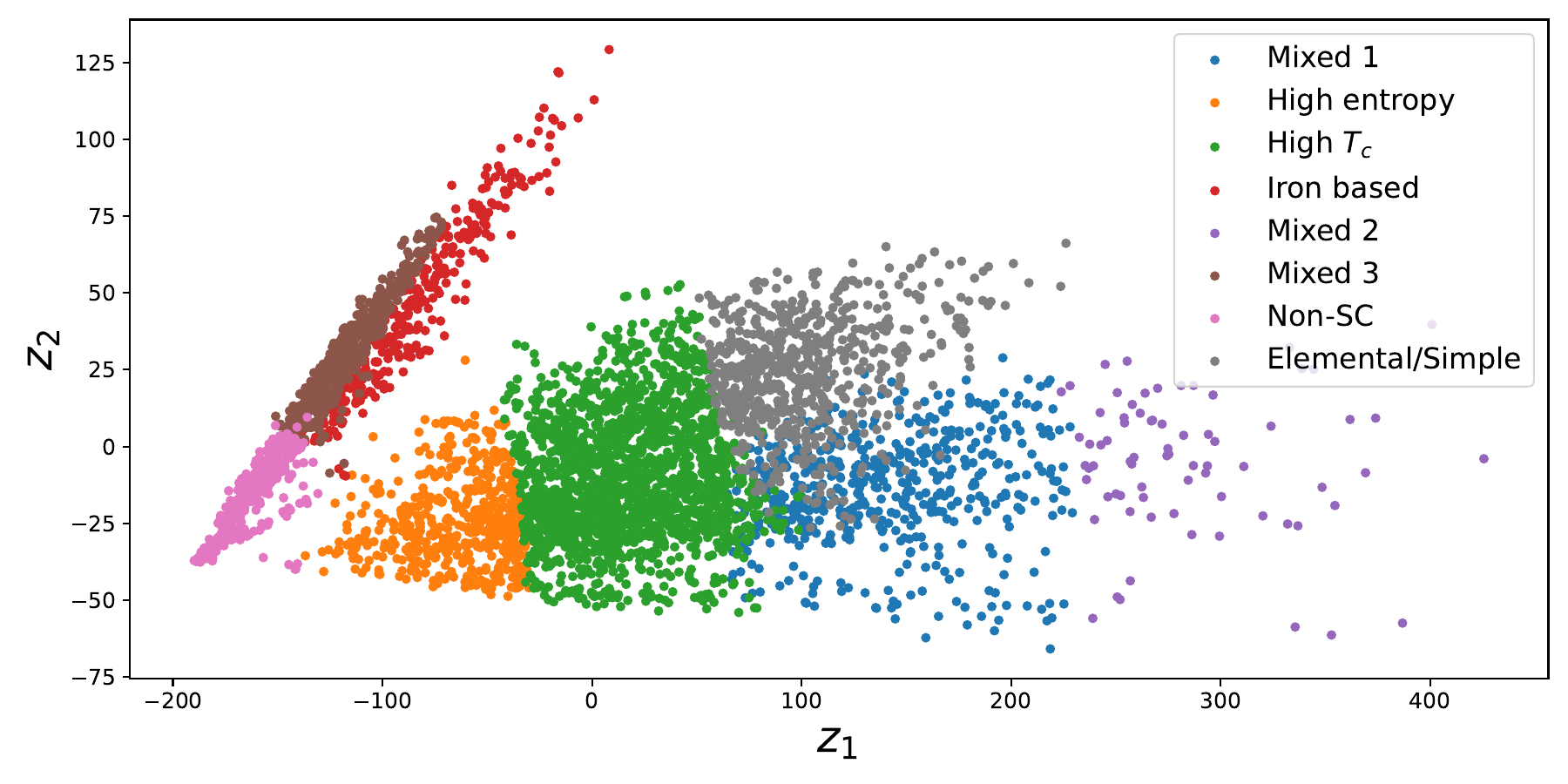}}
    \caption{
     Low-dimensional embedding of superconducting materials
obtained from classical multidimensional scaling (MDS)
applied to trained NTK distances. Colors correspond to
clusters identified through unsupervised k-means clustering
in the embedded representation space. Several clusters align
with known superconducting families, including cuprates,
iron-based superconductors, undoped or otherwise superconductors with ``simple" chemical formulae,
and non-superconducting compounds. As this information is not present during training, we manually assign labels to the generated clusters; each cluster taxonomy is determined by the the first 200 compounds, measured by distance to the centroid.}
    \label{Main_result_cluster}
  \end{center}
\end{figure*}

We apply this framework to superconducting materials data based on the SuperCon dataset \cite{supercon_nims}. We test several cluster sizes, searching for maximum explained variance, for the smallest number of clusters (see Appendix). We find that $N=8$ gives us the best uniform covering of the data. The resultant clusters are presented in Fig.~\ref{Main_result_cluster}, which is one of our main results. These clusters are generated without additional learning or supervision, from the matrix $d(x,x')$ and with the application of the MDS algorithm. 

\begin{table}[htbp]
    \centering
    \caption{Mean Silhouette Scores per Cluster}
    \label{tab:silhouette_scores}
    \begin{tabular}{@{} c c @{}} 
        \toprule
        \textbf{Cluster} & \textbf{Mean Silhouette} \\
        \midrule
        Mixed 1 & 0.175 \\
        High Entropy & 0.415 \\
        High $T_c$ & 0.450 \\
        Iron Based & 0.350 \\
        Mixed 2 & 0.200 \\
        Mixed 3 & 0.140 \\
        Elemental/Simple & 0.300 \\
        Non-SC & 0.425 \\
        \bottomrule
    \end{tabular}
\end{table}

In addition, some clusters are difficult to label and do not contain an obvious preponderance from a specific chemical family. To check the reliability of this clustering, we inspect their relative silhouettes \cite{rousseeuw1987silhouettes}. Inspecting Tab~\ref{tab:silhouette_scores}, we see that the clusters with higher silhouette scores correspond to well-known superconducting material classes as well as the non-superconducting class. This is an indication that the structure of the latent space has learned some information about the underlying physics.

\subsection{Notion of Neighborhoods}

To answer the question concerning the significance of distances presented by $d(x_i,x_j)$, we note that the Shepard plots of Fig.~\ref{fig:shepard} offer a valuable observation: objects in the embedding space are often closer than the metric related to their input space, i.e., $d_{\textrm{euclidean}} = x_i \cdot x_j$. To inspect this, we explicitly investigate the NTK distance, \emph{centered} on select compounds. In order to normalize and define a bona fide distance metric, we normalize the NTK distance $d(x_i,x_j)$ by the diagonal of the NTK Gram matrix, i.e., $d_{ij}^2 \to \frac{d_{ij}^2}{K(x_i,x_i)+K(x_j,x_j)}$.
Then, to illustrate the geometric structure of the space spanned by $d_{ij}$, which we refer to as ``material space", we take two archetypal compounds as origin and order the materials around it by their radial distance. This is akin to measuring pair correlation functions -- a familiar practice in the physics of materials. \ce{Y1Ba_{1.92}Ca_{0.074}Cu4O8} is a famous high temperature superconductor with $T_c \approx$  80 K, and \ce{CeCoIn5}
a heavy fermion superconductor with $T_c$ ~ 1.8  K. 

\begin{figure*}[!t]
    \centerline{\includegraphics[width=\textwidth]{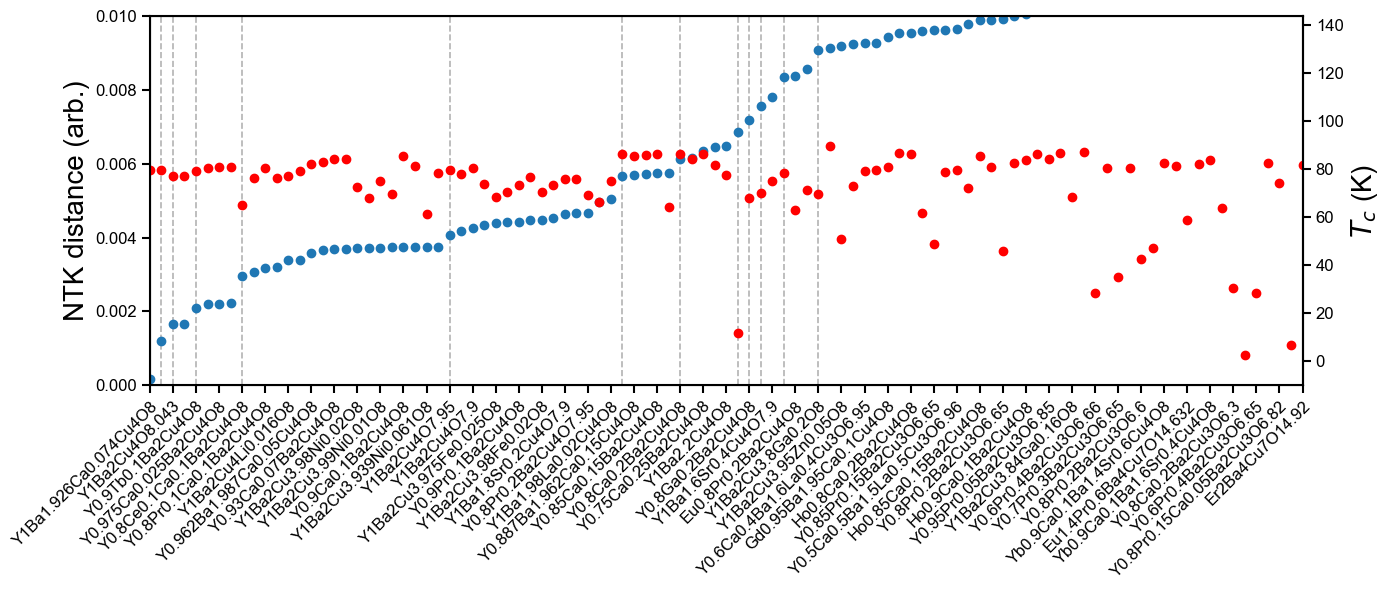}}
    \centerline{\includegraphics[width=\textwidth]{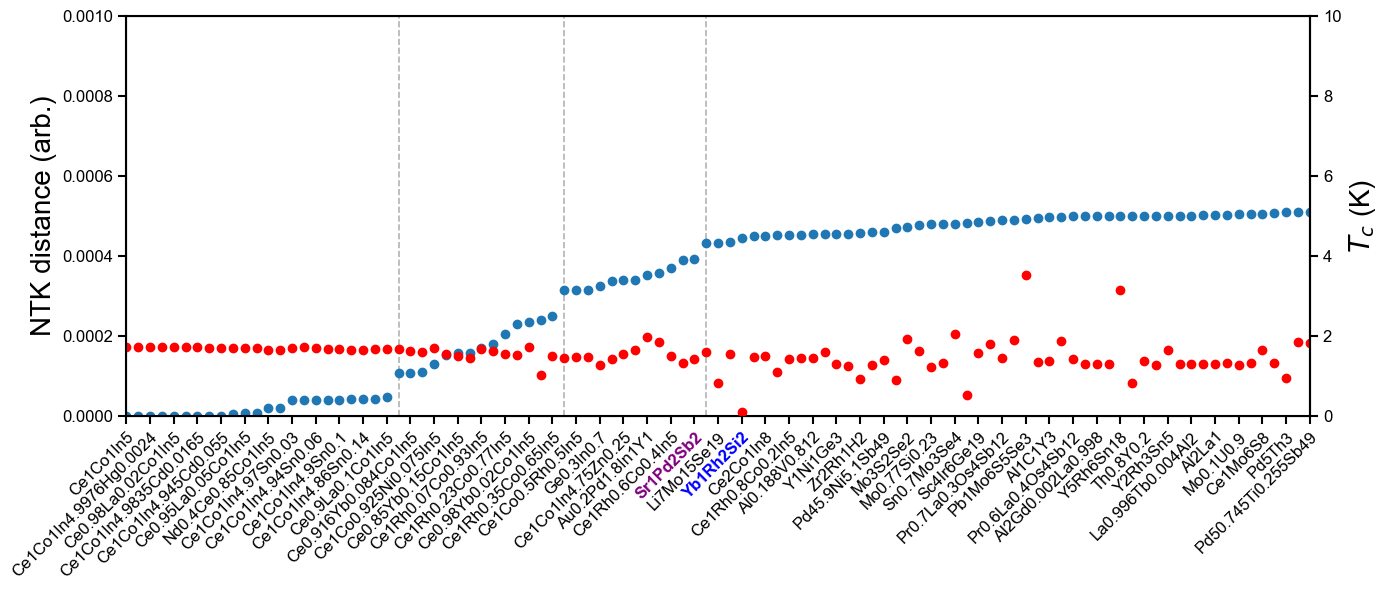}}
    \caption{
     Neighborhood structure induced by the trained NTK geometry.
Materials are ordered according to radial distance from
representative compounds in the learned NTK metric
(top: YBa$_2$Cu$_4$O$_8$-based cuprate; bottom:
CeCoIn$_5$ heavy-fermion superconductor). Blue points
show NTK distance, while red points indicate the superconducting
transition temperature $T_c$. Gray dashed lines denote
distance gaps used to identify neighborhood shells in the
learned geometry. Nearby compounds frequently share
chemical motifs, doping structure, or correlated-electron
phenomenology despite substantial variation in $T_c$,
indicating that the trained NTK captures transferable
physical structure beyond the original supervised learning
task.
    }
    \label{distances}
\end{figure*}

In Fig.~\ref{distances} (top) we show data for YBCO, and Fig.~\ref{distances}(bottom) describes CeCoIn$_5$. We observe several notable trends. Relative to the same distance, we plot $T_c$ against the NTK distance. Remarkably, the distance appears uncorrelated with the variation in $T_c$. Compounds with $d \to 0$ nevertheless show fluctuations of $\Delta T_c \approx \pm 20K$. This astounding result is explained in part by the data itself: the NTK distance measure correctly identified doping, chemical similarity, and the structures of generic chemical formula classes (chemical families, such as the typical $1 2 3 7$, $1 1 3$, etc). 

Fig.~\ref{distances} (bottom), we reveal an even sharper distinction between inputs, which is once more uncorrelated with any label in the supervised learning.
To properly model ``neighborhoods", we introduce a notion of gaps in the distance: we separate regions based on $d > d_c$, which is identified individually for every compound family. These gaps are plotted with a gray dashed line in Fig.~\ref{distances}. 

Initially (within the nearest-neighbor) for CeCoIn$_5$, we find mostly doped versions of the same basic unit (i.e., e.g. Ce$_{0.98}$La$_{0.04} \ldots$). However, within the next-nearest neighbor cluster, we discover a cousin of CeCoIn$_5$, the heavy fermion and strange metal system YbRh$_2$Si$_2$. The learned neighborhood connects $\mathrm{CeCoIn_5}$ with $\mathrm{YbRh_2Si_2}$ although their compositional vectors have zero overlap because they share no constituent elements. Neither heavy-fermion behavior nor strange-metal behavior was supplied as a training label.

The figure demonstrates that the network has learned features significantly different from the task it was trained on, $T_c$.  
Materials with very different superconducting $T_c$, but with other similarities such as similar crystal structure or generic chemical formula, or fundamental physical properties, such as strange metallicity or heavy-fermion ground states, are remarkably close by thanks to the NTK distance.

\subsection{Measuring uncertainty}
So far, our distance measure $d_{ij}$ avoided the element $K(x_i,x_i)$, since $d_{ii} = 0$ by definition. Here we give a direct connection to the object $K(x_i,x_i)$ and discuss its implications on the appearance of clusters in our trained network.

To connect this quantity to uncertainty, consider small, zero-mean
parameter fluctuations with covariance
$\langle\delta\theta_\mu\delta\theta_\nu\rangle
=\sigma_\theta^2\delta_{\mu\nu}$.
Within the tangent approximation, the resulting output fluctuation is
$\delta f(x)=\nabla_\theta f_{\theta_*}(x)\cdot\delta\theta$; this can thought of as a connected two-point function, which therefore satisfies
$G_{\mathrm c}(x,x')
=\langle\delta f(x)\delta f(x')\rangle
=\sigma_\theta^2 K(x,x')$~\cite{Immer2021Linearization}.
The coincident-point limit gives
$\operatorname{Var}[f(x)]
=G_{\mathrm c}(x,x)
=\sigma_\theta^2K(x,x)$,
directly paralleling the identification of local fluctuations with
the diagonal of a connected two-point function in statistical
field theory~\cite{Kardar2007Fields}.
Accordingly, $K(x,x)$ measures the output variance per unit isotropic
parameter variance, while the associated uncertainty scale is
$\sigma_f(x)=\sigma_\theta\sqrt{K(x,x)}$.
Larger diagonal values thus identify inputs whose predictions
fluctuate more strongly under the same level of uncertainty in the
learned parameters. This interpretation applies locally around the
trained configuration, requires no infinite-width assumption, and
is closely related to uncertainty constructions based on
linearized networks and trained empirical
NTKs~\cite{Wilson2025EmpiricalNTK}.
\begin{figure*}[!t]
    \centerline{\includegraphics[width=\textwidth]{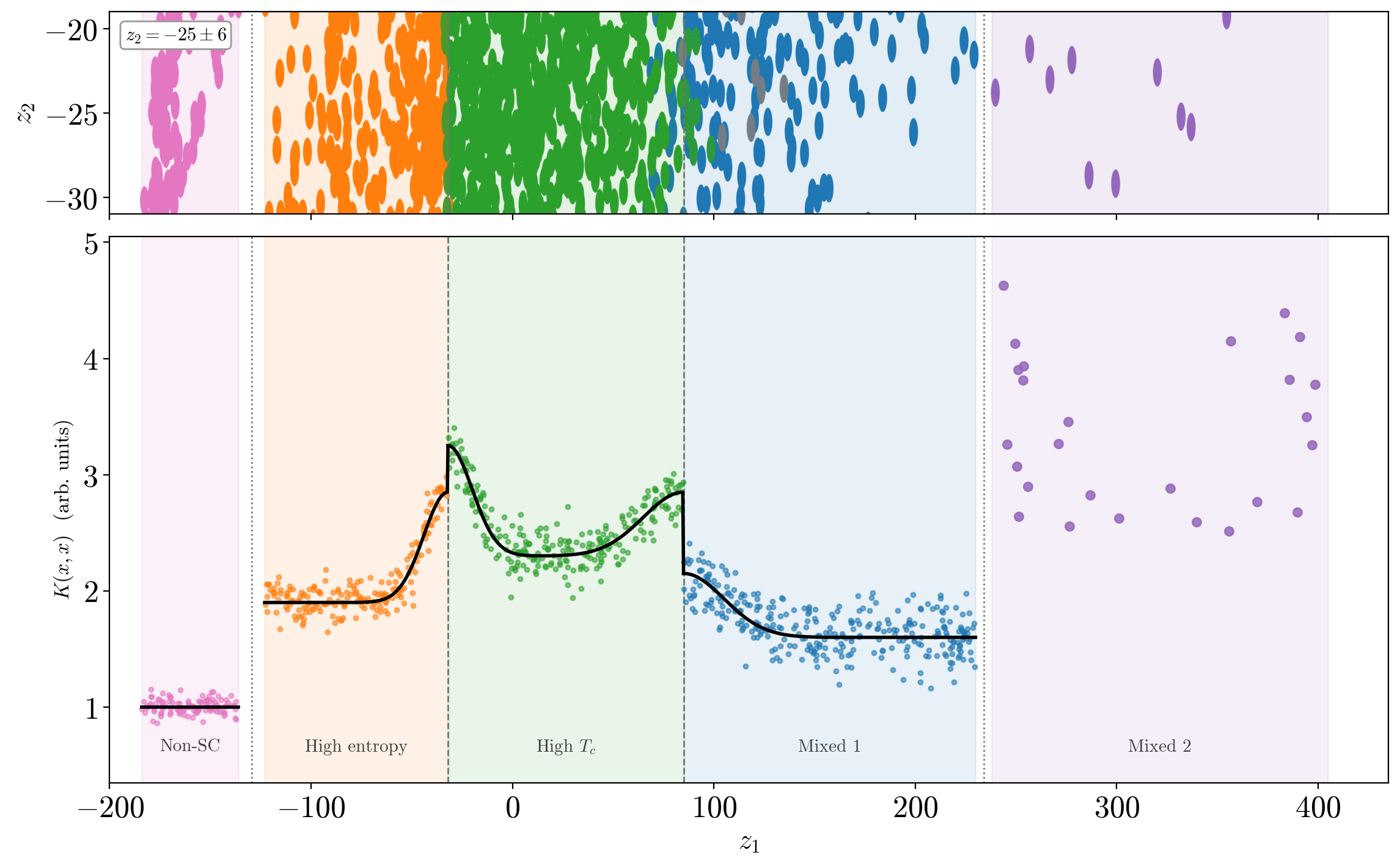}}
    \caption{
     Measure of $K(x,x)$ as uncertainty in clustering. We take a cut along a fixed $z_2$ value $z_2 = -25$ and average over a window of $\pm6$ in that range. We then traverse $z_1$. We find a correlation between boundaries of the identified clusters and increases in $K(x,x)$. In areas of low explainability/silhouette, such corresponding changes in $K(x,x)$ were not observed.  
    }
    \label{NTK_cut}
\end{figure*}
In Fig.~\ref{NTK_cut}, we show a particular application to this uncertainty measure. We fix $z_2 = -25$ and average over a window of $\pm 6$. In turn, we plot $K(x,x)$ across the coordinate $z_1$. We find an increase in $K(x,x)$ precisely at the boundaries of well-identified clusters, and in areas where different samples from different clusters leak into each other. Such an identifiable behavior was not established in areas of poor explanability and correspondingly low silhouette.
\section{Discussion}

In this work, we have shown that the trained neural tangent kernel provides an interpretable probe of the representation learned by a supervised neural network. Using superconducting materials as a model system, we demonstrated that the trained NTK induces a low-dimensional geometry that faithfully reproduces the predictive capability of the original network while organizing materials according to physically meaningful relationships. Unlike the NTK at random initialization, the trained kernel substantially reorganizes the geometry inherited from the input representation, revealing neighborhoods, clusters, and dominant directions associated with the learned task. Importantly, these structures are recovered without retraining the network or introducing additional representation-learning objectives.

More broadly, our work complements the long tradition of representation learning in machine learning. Classical approaches such as independent component analysis, sparse coding, and more recent self-supervised methods explicitly optimize latent representations from data. Our work addresses the complementary problem of representation recovery rather than representation learning. Once a supervised predictor has been successfully trained, we ask what scientific representation it has already acquired and how it can be interpreted. Our results suggest that the trained NTK provides a practical means of recovering this representation without auxiliary objectives or additional optimization.

Perhaps the most surprising aspect of our results is that this representation emerges from elemental composition alone. As we stated in the introduction, popular approaches to generate a meaningful low-dimensional manifold require an abundance of features fed during training. In contrast, the predictor considered here receives only elemental composition of materials. The emergence of a meaningful representation under these conditions suggests that elemental composition itself contains a surprisingly rich latent organization. Supervised learning then uncovers this organization, and the trained NTK provides a means of recovering and interpreting it.

More broadly, our results suggest that trained tangent kernels should be viewed not only as analytical tools for understanding optimization, but also as scientific objects in their own right. Their induced geometry provides an interpretable representation of learned scientific structure, offering a bridge between modern representation learning, kernel methods, and the analysis of complex materials datasets.

\begin{acknowledgements}
We thank D. Chowdhury, P. de Faira, P. Ginsparg, E. Kim, O. Lesser and T. Scaffidi for useful discussions. 
D.K. acknowledges the support of the Abrahams postdoctoral fellowship of the Center for Materials Theory, Rutgers University and the Zuckerman STEM Fellowship.
This work as completed in part at Aspen Center for Physics, which is supported by National Science Foundation grant PHY-2210452.
\\
*
\end{acknowledgements}
\bibliography{main.bbl}
\end{document}